\documentclass[a4paper, superscriptaddress,groupedaddress,nofootnoteinbib,11pt]{article}  % for review and submission
\usepackage{amssymb,amsmath,graphicx,color,microtype}
\usepackage{varioref}
\usepackage{graphicx}
\usepackage{txfonts}
\usepackage{jcappub}
\usepackage{epsf}
\usepackage{multirow}
\usepackage{epstopdf}
\usepackage {amssymb}
\usepackage{tcolorbox}
\usepackage{tabularx}
\usepackage{array}
\usepackage{colortbl}
\tcbuselibrary{skins}
\usepackage{booktabs}

\newcolumntype{Y}{>{\raggedleft\arraybackslash}X}
\newcommand{\nc}{\newcommand}
\nc{\ba}{\begin{eqnarray}}
\nc{\ea}{\end{eqnarray}}
\newcommand\be{\begin{equation}}
\newcommand\ee{\end{equation}}

\usepackage{tabularx}
\tcbset{tab2/.style={enhanced,fonttitle=\bfseries,fontupper=\normalsize\sffamily,
colback=yellow!10!white,colframe=red!50!black,colbacktitle=Salmon!40!white,
coltitle=black,center title}}

\nc{\e}{{\bf{e}}}
\nc{\kk}{{\bf{k}}}
\nc{\pp}{{\bf{p}}}

\nc{\bfk}{{\bf{k}}}
\nc{\bfx}{{\bf{x}}}
\nc{\bfp}{{\bf{p}}}

\nc{\eH}{{\epsilon_H}}
\nc{\calP}{{\cal P}}
\nc{\im}{{ \mathrm{Im} } }

\begin{document}
	
\title{Observational Constraints on the Primordial Curvature Power Spectrum}

\author{Razieh Emami$^{1}$ and George F. Smoot$^{2,3,4}$}

\affiliation
    { $^1$ Institute for Advanced Study, The Hong Kong University of Science and Technology, Clear Water Bay, Kowloon, Hong Kong; iasraziehm@ust.hk \\
	$^2$ Helmut and Anna Pao Sohmen Professor-at-Large, IAS, Hong Kong University of Science and Technology,
	Clear Water Bay, Kowloon, 999077 Hong Kong, China \\
	$^3$ Paris Centre for Cosmological Physics, APC, AstroParticule et Cosmologie, Universit\'{e} Paris Diderot,
	CNRS/IN2P3, CEA/lrfu, %Observatoire de Paris,  
	Universit\'{e} Sorbonne Paris Cit\'{e}, 10, rue Alice Domon et Leonie Duquet,
	75205 Paris CEDEX 13, France;  \\
	$^4$
	Physics Department and Lawrence Berkeley National Laboratory, University of California, Berkeley,
	94720 CA, USA; gfsmoot@lbl.gov
	}

\abstract{\\
CMB temperature fluctuation observations provide a precise measurement of the primordial power spectrum on large scales, corresponding to wavenumbers $10^{-3}$ Mpc$^{-1} \lesssim  k  \lesssim  0.1 $ Mpc$^{-1}$, \cite{Pearson:2002tr, Reichardt:2008ay, Brown:2009uy, Larson:2010gs, Hlozek:2011pc, Aslanyan:2014mqa, Ade:2015lrj, Aghanim:2015xee}.
Luminous red galaxies and galaxy clusters probe the matter power spectrum on overlapping scales 
(0.02 Mpc$^{-1} \lesssim  k  \lesssim  0.7$ Mpc$^{-1}$; \cite{Reid:2009xm, Vikhlinin:2008ym, Tinker:2011pv, Sehgal:2010ca, Reddick:2013cha, Bleem:2014iim, Gil-Marin:2015nqa, Beutler:2016ixs, Ravenni:2016vjd, Armengaud:2017nkf}), 
while the Lyman-alpha forest reaches slightly smaller scales (0.3 Mpc$^{-1} \lesssim  ~ k  \lesssim 3$ Mpc$^{-1}$; \cite{McDonald:2004eu}). 
These observations indicate that the primordial power spectrum is nearly scale-invariant with an amplitude close to $2 \times 10^{-9}$, \cite{Tegmark:2002cy, Nicholson:2009pi, Komatsu:2010fb, Dunkley:2010ge, Keisler:2011aw, Hlozek:2011pc, Bird:2010mp}.  
These observations strongly support Inflation and motivate us to obtain observations and constraints reaching to smaller scales on the primordial curvature power spectrum and by implication on Inflation.
We are able to obtain limits to much higher values of $k \lesssim 10^5$ Mpc$^{-1}$ 
and with less sensitivity even higher $k \lesssim 10^{19}- 10^{23}$ Mpc$^{-1}$ 
using limits from CMB spectral distortions and other limits on ultracompact minihalo objects (UCMHs) and Primordial Black Holes (PBHs).
PBHs are one of the known candidates for the Dark Matter (DM). 
Due to their very early formation, they could give us valuable information about the primordial curvature perturbations. 
These are complementary to other cosmological bounds on the amplitude of the primordial fluctuations. 
In this paper, we revisit and collect all the published constraints on both PBHs and UCMHs. 
We show that unless one uses the CMB spectral distortion, 
PBHs give us a very relaxed bounds on the primordial curvature perturbations. 
UCMHs, on the other hand, are very informative over a reasonable $k$  range ($3 \lesssim k \lesssim 10^6$ Mpc$^{-1}$) and 
lead to significant  upper-bounds on the curvature spectrum. 
We review the conditions under which the tighter constraints on the UCMHs could imply extremely strong bounds on the fraction of DM that could be PBHs in reasonable models. 
Failure to satisfy these conditions would lead to over production of the UCMHs which is inconsistent with the observations. Therefore, we can almost rule out PBH within their overlap scales with the UCMHs. 
We compare the UCMH bounds coming from those experiments which are sensitive to the nature of the DM, 
such as $\gamma$-rays, Neutrinos and Reionization, 
with those which are insensitive to the type of the DM, e.g. the pulsar-timing as well as CMB spectral distortion. 
We explicitly show that they lead to comparable results which are independent of the type of DM. 
These bounds  however  do depend on the required initial density perturbation, 
i.e. $\delta_{min}$. It could be either a constant or a scale-dependent function. 
As we will show, the constraints differ by three orders of magnitude depend on our choice of required initial perturbations.}
	
\maketitle 

%%%%%%%%%%%%%%%%%%%%%%%%%%%%%%%%%%%%%%%%%%%%%%%%%%%%%
\section{Introduction}
One of the main issues in modern cosmology is the existence for and the nature of a non-baryonic missing matter, called the dark matter (DM). 
Different observations continue to show the necessity of DM at various scales and redshifts. 
However, the nature of this component is still not clear. 
There are plenty of different candidates to describe the DM, which could be categorized into various classes. 
One (very well-known) class of models contain particles with masses range out from (very) light scalars such as axions, \cite{axion}, to heavier particles like the neutralino, 
\cite{Jungman:1995df}, or even ultra-massive particles like WIMPs, \cite{Kolb:1998ki}. 
Another class of models is using astrophysical compact objects as the DM. 
This contains the primordial black holes (PBHs) as well as ultracompact minihalos (UCMHs). \\
The idea of PBH was first proposed by \textit{Zel'dovich} and \textit{Novikov},\cite{zeldovich}. 
And, it was further developed by Hawking, \cite{Hawking:1971ei, Hawking:1974rv}. 
Chapline was the first person who put forward the idea of using the PBHs as DM, \cite{Chapline}. 
The discovery of gravitational waves by LIGO team, \cite{Abbott:2016blz}, from two merging black holes with masses of the order $30 M_{\odot}$, revived this idea in \cite{Bird:2016dcv, Sasaki:2016jop}. 
This brought back the previous interest on using PBHs as a candidate for the DM by a lot of authors, \cite{Cholis:2016kqi, Gaggero:2016dpq, Ali-Haimoud:2016mbv, Nakama:2016gzw, Clark:2016nst, Schutz:2016khr, Garcia-Bellido:2017mdw, Domcke:2017fix}. There have been also many tests to probe these early universe scenarios, \cite{Namikawa:2016edr, Kawasaki:2016pql,Carr:2016drx,
Kuhnel:2017pwq, Akrami:2016vrq, Chen:2016pud, Clesse:2016ajp, Nakamura:2016hna, Raccanelli:2016cud, Brandt:2016aco, Munoz:2016tmg, Inoue:2017csr, Chiba:2017rvs, Gong:2017sie, Fuller:2017uyd, Georg:2017mqk, Garcia-Bellido:2017fdg, Rice:2017avg, Dolgov:2017nmh, Belotsky:2017vsr, Kannike:2017bxn, Carr:2017jsz, Ezquiaga:2017fvi, Kovetz:2017rvv}.\\ 
The idea of UCMHs was firstly proposed by Ricotti and Gould, \cite{Ricotti:2009bs} where they put forward the idea of existence of UCMH as a new type of massive compact halo objects, here after MACHO. 
UCMHs were further considered in \cite{Choi:2015qdu, Aslanyan:2015hmi, Beck:2016gkv, Yang:2016cxm, Clark:2016pgn, Anthonisen:2015tda, Kohri:2014lza, Zheng:2014tta, Berezinsky:2013fxa, Yang:2011eg}.\\
One of the biggest differences between the PBHs and the UCMHs is the time of their formation. 
Due to the necessity of having a much larger  primordial density perturbations to form PBHs, the gravitational collapse would happen much earlier, during the radiation dominance for the PBH and not until the matter-radiation equality for the UCMHs. \\

Before going through the detail description of PBHs and the UCMHs, it is worth briefly discussing some possible formation mechanisms. As the requirement for creating PBHs are more severe as compared with the UCMHs, we just focus on their formation. For sure, some of these mechanisms could be also applied for the case of the UCMHs as well. 
The first mechanism to produce PBHs was proposed in 1993 by Dolgov and Silk, \cite{Dolgov:1992pu}, where they use the QCD transition to form PBH with mass of the order of the solar mass. This interesting idea was later developed in \cite{Jedamzik:1996mr}. Although further achievements in the QCD theory disfavored the first order phase transition, which is required for this model to work, it was a good starting work for people to think about different formalisms that could possibly create PBHs. In principle, these mechanisms could be divided into few different categories. 
In the following, we briefly mention them. We reference the interested reader into the papers for further study. \\
(1) Inflationary driven PBHs: One of the main candidates to produce the PBHs is using inflation. However, there are some issues that must be considered in this regard. In the single field inflationary models, the density fluctuations are in general well below than the required threshold to generate the PBH. So modifications to the inflaton potential are indeed necessary to achieve the minimal requirements for the PBH production. This can be done either by considering the blue tilted potentials or some forms of the running of the spectral index. However, the generated mass is too tiny, well below the solar mass. There are several mechanisms that can be useful to boost the mass range of the PBHs into the astrophysical or the cosmological values. For example, this can be done in different multi-field inflationary models, such as in Hybrid inflation, \cite{GarciaBellido:1996qt, Kawasaki:2015ppx, Clesse:2015wea}, double inflation, \cite{Yokoyama:1998pt, Kawasaki:1998vx, Kawasaki:1997ju, Kawaguchi:2007fz, Frampton:2010sw}, curvaton inflation, \cite{Kawasaki:2012wr, Kohri:2012yw, Bugaev:2012ai}, in particle production during Inflation, \cite{Linde:2012bt, Erfani:2015rqv, Garcia-Bellido:2016dkw}, or during trapped inflation, \cite{Cheng:2016qzb}, and inflection point inflation, \cite{Garcia-Bellido:2017mdw}. In these models, the small scale perturbations could be boosted into scales ranging up from the stellar mass PBH to a super-massive PBHs etc.  \\
(2) Enhancement of PBH due to secondary effects: Regardless of the origin of the primordial inhomogeneities, there could be some secondary effects which somehow make an enhancement in the formation of the PBHs. Such an improvement may happen during a sudden decrease in the pressure, e.g. when the universe does pass through a dust-like phase happening as a results of the existence of Non-relativistic particles, \cite{Khlopov:1980mg, Khlopov:1985jw}, or during the reheating phase, \cite{Carr:1994ar}. \\
(3) PBH formation without initial inhomogeneities: As another possibility, PBHs could also be formed during some sorts of the phase transitions in the universe. The very important point is that in this case, they do not even need any initial inhomogeneities to start with. This could happen from bubble collision, \cite{Crawford:1982yz, Kodama:1982sf, Hawking:1982ga, La:1989st, Moss:1994iq, Konoplich:1999qq}
, from the collapse of the cosmic string, \cite{Hogan:1984zb, Polnarev:1988dh, Hawking:1987bn, Garriga:1993gj, Caldwell:1995fu, Cheng:1996du, Hansen:1999su, MacGibbon:1997pu}, or from the domain walls, \cite{Caldwell:1996pt, Berezin:1982ur, Rubin:2000dq, Khlopov:2000js}. Generally cosmic strings would be anticipated to be at a smaller mass range, \cite{Garriga:1993gj, MacGibbon:1997pu}.\\
There are many constraints on the abundance of both of PBHs as well as the UCMHs. 
And, as we point out in what follows, it turns out that they can not make the whole of the DM. \\
Since the compact objects are formed from density fluctuations with (large) initial amplitudes, the above constraints on their initial abundance can be translated back to constraints on the primordial curvature power spectrum. In addition, due to the large available mass range for them, these limits could be extended into very small scales which are completely out of the reach of cosmological constraints which are coming from the combination of CMB, large scale structure and Lyman-$\alpha$ and covers the scales ($ k \simeq 10^{-4} - 1 Mpc^{-1}$). Indeed these constraints could be go down to  $k \simeq 10^{7} Mpc^{-1}$ for UCMHs while extend much more to $k \simeq 10^{23} Mpc^{-1}$ for the PBH. 
So it would be extremely useful to increase our information about the very-much smaller scales that are otherwise to be probed by other kind of experiments. \\
In this article, we will use and update the known limits on the initial abundance of PBH as well as the UCMHs from a combination of many different kind of experiments, both current as well as considering futuristic ones. 
Assuming a Gaussian shape for the initial fluctuations, we will then connect these limits into the constraints on the amplitude of the initial curvature perturbations. 
For the case of PBHs, (in general) the constraints are very relaxed, $\mathcal{P}_{R}(k) \simeq 10^{-2}$, in their very wide $k$-space window. However, they could be more constrained if we use the CMB spectral distortions (CMB SD) as a probe. CMB SD increase the constraints by several orders of magnitudes, $\mathcal{P}_{R}(k) \simeq 10^{-5}$ for COBE/FIRAS and $\mathcal{P}_{R}(k) \simeq 10^{-8}$ for the proposed PIXIE. 
However, it is worth mentioning that this probe is only sensitive to $ (1 \leqslant \left(k/Mpc^{-1}\right) \leqslant 10^{4})$. 
The constraints coming from the UCMHs, on the other hand, are more informative. As we will see, there are several different probes on this candidate coming either from the particle physics side,like the Gamma-Ray, Reionization or Neutrinos, or from the gravity side, from Pulsar Timing. Some of these tests put more constraints on the initial amplitude of the curvature perturbation while the others, like for example the Reionization test, are more relaxed. In addition, as we will see in the following, there is an ambiguity in the lower bound of density perturbation to build up the UCMHs. Some authors considered it to be a constant of order $\delta \simeq 10^{-3}$ while the others have taken its scale dependent into account. As well will see relaxing such an assumption would lead to few orders of magnitude changes in the constraints on the primordial power spectrum. Finally, we could also consider the constraints coming from the CMB SD for the UCMHs as well. Interestingly, their upper-bounds on the initial curvature power-spectrum is of the same order as the bounds on the PBHs. \\
The rest of the papers is entitled as the follows. \\
In Sec. \ref{PBH-introduction}, we introduce PBH as well as their inferred observational constraints on the initial curvature power-spectrum. In Sec. \ref{UCMH-introduction}, we first introduce UCMH and then would list up the whole constraints on the amplitude of the primordial spectrum. In Sec. \ref{cmb-lss-lya-constraints} we will present the constraints on the spectrum coming from the combination of CMB, LSS and the Ly-$\alpha$. We conclude in Sec. \ref{conclusion}.

\section{A consideration of PBHs and their limits on the curvature power spectrum}
\label{PBH-introduction}
As it is well-known, one of the examples of the astrophysical compact objects is the primordial black holes, here after PBHs, which are assumed to be created in the early stage of the universe and from the density perturbations. The intuitive picture behind their formation is the following: suppose that the density perturbations at the stage of the horizon re-entry exceeds a threshold value, which is of order one ($\gtrsim 0.3$). Then the gravity on that region would overcome the repulsive pressure and that area would be subjected to collapse and will form PBH. Such a PBHs would span a very wide range of the mass range as follows, \cite{Carr:2009jm}
\ba
\label{mass-range-PBH}
M \simeq \frac{c^3 t}{G} \simeq 10^{15} \left(\frac{t}{10^{-23}}\right) g 
\ea
where $t$ denotes the cosmic time. \\
Eq. (\ref{mass-range-PBH}) gives us an intuitive way to better see the huge range of the mass for the PBHs. Furthermore, we could even see that earlier formation of the PBHs lead to very tiny mass range. For example around the Planck time, $t \simeq 10^{-43}$, we could create PBHs of the order $10^{-5}$ g. Going forward in time, we could also see that around $t \simeq 10^{-5}$s PBHs as massive as sun could be produced this way. Once they are created, they evaporate through the Hawking radiation and with the following lifetime, 
\ba
\label{life-time-PBHs}
\tau(M) &\sim& \frac{G^2 M^3}{\hbar c^4} \sim 10^{64} \left(\frac{M}{M_{\odot}}\right)^3 yr
\ea
here $\tau(M)$ refers to the lifetime of the PBHs. A first look at Eq. (\ref{life-time-PBHs}) does show that PBHs with a mass less than $\sim 10^{15}$g  or about $10^{-18}$ M$_\odot$ would have evaporated by now. However, those with larger mass could still exist by today and they could be used as a candidate for the DM. 
\\
There are very tight observational constraints on the (initial) abundance of the PBHs. 
The purpose of this section is to use these constraints and translate them back to the primordial curvature power-spectrum constraints, as was first pointed out in \cite{Josan:2009qn}. 
In our following analysis, we would consider the full available range of the scales from $(10^{-2}-10^{23}) Mpc^{-1}$, 
therefore extending their case study to few more orders of magnitude. 
In order to get a feeling about the above wide range of the scales, it is worth to assume that at every epoch, the mass of the PBH is a fixed fraction $f_M$ of the horizon mass, i.e. $M_{H}\sim \frac{4\pi}{3} \left( \rho_r H^{-3}\right)_{k = aH}$. 
We assume $f_M \simeq \left(1/3\right)^{3/2}$. It is then straightforward to find, 
\ba
\label{mass-range}
\left(\frac{M}{M_{eq}}\right) \sim \left(\frac{g_{eq}}{g}\right)^{1/3} \left(\frac{k_{eq}}{k}\right)^2
\ea
where $M_{eq} = 1.3 \times 10^{49} \left(\Omega^0_m h^2\right)^{-2}$g,  $g_{eq} \simeq 3$ and $k_{eq} = 0.07~\Omega^0_m h^2 Mpc^{-1}$. 
We would also present few futuristic constraints on both of the curvature perturbation as well as the abundance. 
The type of the constraints that we get for the amplitude of the curvature perturbation is much relaxed as compared with similar constraints coming from the CMB and the LSS. However, at the same time, we go much beyond their scale. Therefore, this study gives us  bounds on the inflationary models not otherwise tested. \\
Having presented a summary about the PBHs, we now investigate the observational constraints on the abundance of the PBHs. For this purpose, we define two important parameters, \\
\textit{\textbf{ (a) Initial abundance of PBHs or the mass fraction:}} 
\ba
\beta(M_{PBH}) \equiv \left(\frac{\rho^{i}_{PBH}}{\rho^{i}_{crit}}\right)
\ea
where the index $i$ denotes the initial value. 
This quantity is particularly important for $M_{PBH} < 10^{15}$g though it is also widely used for the other branch of the PBH masses. And, can be used when we would like to put constraints on the initial curvature power-spectrum, as we will see in the following. \\
\textit{\textbf{ (b) (Current) fraction of the mass of Milky Way halo in PBHs:}} 
\ba
\label{halo-fraction-mass}
f_{h} \equiv \left(\frac{M^{MW}_{PBH}}{M^{MW}_{CDM}}\right)\approx \frac{\rho^{0}_{PBH}}{\rho^{0}_{CDM}} \approx 5 \Omega^0_{PBH}
\ea
This parameter is very useful for PBHs with $M_{PBH}> 10^{15}$g. In this case, PBHs could be interpreted as a candidate for the CDM. \\
Finally it is worth expressing the relationship between the above two quantities, 
\ba
\label{beta-fh}
f_{h} = 4.11 \times 10^{8} \left(\frac{M_{PBH}}{M_{\odot}}\right)^{-1/2} \left(\frac{g_{\star,i}}{106.75}\right)^{-1/4}\beta(M_{PBH})
\ea
In the following subsections, we first point out different observational constraints on the initial abundance of the PBH and then would present a consistent way to read off the constraints on the initial curvature power spectrum. 
\subsection{Observational constraints on the abundance of the PBHs}
\label{PBH-Experiments}
Below we present different observational constraints on the abundance of the PBHs including both of the current as well as the futuristic constraints. We update and extend up the previous analysis by \cite{Josan:2009qn} in few directions. 
	
\subsubsection{Disk heating }
As was pointed out in \cite{Josan:2009qn, Lacey, carr:1999}, as soon as a massive halo object traverse the Galactic disk, 
it would heat up the disk and will increase the velocity dispersion of the disk stars. Finally, it would put a limit on the halo fraction in the massive objects. Though these limits are coming from the consideration of compact objects, we would expect them to be almost the case for the primordial black holes as well.
The relevant limit is presented in Table (\ref{tab:PBH}).

\subsubsection{Wide binary disruption}
Massive compact objects within the range $ 10^{3} < (M_{PBH}/M_{\odot}) < 10^{8} $ would affect the orbital parameters of the wide binaries,\cite{Bahcall, Weinberg:1987, Quinn:2009zg}. This would again  constrains the abundance of the primordial black holes as it is presented in Table (\ref{tab:PBH}).

\subsubsection{Fast Radio Bursts}
As it was recently pointed out in \cite{Munoz:2016tmg}, the strong gravitational lensing of the (extragalactic) Fast Radio Bursts, here after FRB, by PBH would result in a repeated pattern of FRB. The associated time delay for these different images are of the order $\tau_{Delay} \sim \mathcal{O}(1) \left(M_{PBH}/(30 M_{\odot})\right)$ms. On the other hand, the duration of FRB is of the order ms as well. Therefore we should be able to see this repeated pattern out the signal itself for the PBH within the range $10 <M_{PBH}/M_{\odot} < 10^{4}$. There are several ongoing experiments to observe about $10^{4}$ FRB  per year in the near future, such as CHIME. We argue that a null search for such a pattern would constrain the ratio of the Dark Matter in the PBH form to be $\leqslant 0.08 $ for $ M \geqslant  20 M_{\odot}$. The detailed constraints on the $\beta$ is given in Table (\ref{tab:PBH}). It is worth mentioning that the details of the time delay shape does also depend on the PBH mass function; it would be different for the extended as compared with the delta function mass. So in principle one could also play around with that factor too. However, since the whole of the above constraints came for the mass function of the delta form, in order to compare the order of the magnitudes with each other, we avoid considering the extended mass functions for the PBHs. 

\subsubsection{Quasar microlensing}
Compact objects within the mass range $ 10^{-3}< (M_{PBH}/M_{\odot}) < 300 $ would microlens the quasars and amplify the continuum emission though do not alter the line emission significantly, \cite{Dalcanton: 1994,Canizares: 1982}. Non-observation of such an amplification leads the presented limit on the abundance of the primordial black holes as in Table (\ref{tab:PBH}). 

\subsubsection{Microlensing}
Solar and planetary massive compact objects within the Milky Way halo could microlens stars in the Magellanic Clouds, \cite{Paczynski: 1986, Griest: 1991, Alcock:2000ph, Tisserand:2006zx, Alcock:1998fx, Allsman:2000kg}. Indeed, one of the most promising ways to search for the primordial black holes is to look for the lensing effects caused by these compact objects. An experimental upper bound on the observed optical depth due to this lensing is translated into an upper bound on the abundance of the primordial black hole in the way that is presented in Table (\ref{tab:PBH}). 

\subsubsection{GRB femtolensing}
As we already pointed it out above, lensing effects are a very promising approach to constrain the abundance of the PBHs. 
As it is well understood, within some mass ranges the Schwarzschild radius of PBH would be comparable to the photon wavelength. In this cases, lensing caused by PBHs would introduce an interferometry pattern in the energy spectrum which is called the femtolensing. \cite{Gould} was the first person to use this as a way to search for the dark matter objects within the range $ 10^{-16} \lesssim M_{PBH}/M_{\odot} \lesssim 10^{-13}$. A null detection of the femtolensing  puts constraints on the abundance of the PBHs. The limits are shown in Table (\ref{tab:PBH}). 

\subsubsection{Reionization and the $21 cm$ signature}
The recent experimental results, such as WMAP, Planck, etc, have shown that the reionization of the Universe has occurred around $z \sim 6$. However, in the presence of the PBHs, this would be changed. Therefore, there is a limit on the abundance of the PBHs within the range, $ M_{PBH}/ M_{\odot} > 10^{-20}$. 
In addition, as it was shown in \cite{Mack:2008nv}, any increase in the ionization of the intergalactic medium leads to a $21 cm$ signature. In fact, the futuristic observations of $21 cm$ radiation from the high redshift neutral hydrogen could place an important constraint on the PBHs in the mass range $ 10^{-20} \lesssim M_{PBH}/ M_{\odot} \lesssim 10^{-17}$. The limits are shown in  Table (\ref{tab:PBH}).  Keep in mind that these are the potential limits, rather than actual ones. 

\subsubsection{CMB Spectral Distortion}
As PBHs evaporates, they could produce diffuse photons. Indeed the generated photons out the evaporations of the PBHs have two effects. On the one hand they directly induce the $\mu$-distortion at the CMB. And on the other hand, they also heat up the electrons in the surrounding environment. Such electrons later on scatter the photons and produce the $y$-distortion, \cite{Tashiro:2008sf}. Likewise, a PBH absorbing the diffuse material around it would also heat of the environment. the amount of such heating depends upon whether the infall free streams or accretes in a more complicated manner. The upper-limits from the COBE/FIRAS  on CMB spectral distortions then put constraints on the initial abundance of the PBHs. The results are shown in Tab. (\ref{tab:PBH}). Such an upper limit could be pushed down by about three orders of the magnitude if we consider the PIXIE experiments, taking into account that the distortion parameters, especially the $y$-distortion, are proportional to the initial abundance of the PBH, \cite{Tashiro:2008sf}. 

\subsubsection{Photodissociation of Deutrium}
The produced photons by PBH can photodissociate deuterium, D. This leads to a constraint on the abundance of the primordial black holes, \cite{Lindley:1980, Clancy:2003zd, Naselskii: 1978, Mather: 1994, Tashiro:2008sf}, as is presented in Table. (\ref{tab:PBH}).

\subsubsection{Hadron Injection}
PBHs with a mass less than $M_{PBH} < 10^{-24} M_{\odot} $
have enough life time, i.e. $\tau < 10^{3}$s, to affect the light element abundances, \cite{Josan:2009qn, Miyama:1978mp, Kohri:1999ex, Clancy:2003zd} , during the cosmic history of the universe. This leads to a constraint on the abundance of the primordial black holes as it is presented in Table. (\ref{tab:PBH}).

\subsubsection{Quasi stable massive particles}
In any extensions of the standard model of the particle physics, there could be some generations of the stable or long lived(quasi-stable) massive particles, hereafter Quasi-SMP, with the mass of the order $O(100 GeV)$. One way to get these particles produced is through the evaporation of the light PBHs with the mass $M_{PBHs} < 5 \times 10^{-23} M_{\odot}$. Once these particles are produced, they are subject to a later decay, say for example after the BBN, and therefore would change the abundance of the light elements. This leads to a constraint on the abundance of the primordial black holes, \cite{Green:1999yh, Lemoine:2000sq, Khlopov:2004tn}, as is presented in Table. (\ref{tab:PBH}).

\subsubsection{Lightest Supersymmetric particle}
In supersymmetric extension of the standard model, due to the R-parity, the Lightest Sypersymmetric Particle (LSP) must be stable and be one of the candidates for the dark matter. LSP which are being produced through the evaporation of the PBHs, lead to an upper limit on the abundance of the PBHs to make sure not to exceed the observed CDM density, \cite{Josan:2009qn, Carr:2009jm}. The limit is shown in Table.  (\ref{tab:PBH}).

\subsubsection{Planck Mass Relics}
Planck-mass relics could make up the Dark Matter today, if they are stable relics. In order not to exceed the current observed CDM density, there would be an upper limit on the abundance of the primordial black holes, \cite{Carr:1994ar}, as it is presented in Table. (\ref{tab:PBH}). However,  most theorists do believe that physics models have a stable Planck-mass relic. We present both of these options in Fig. \ref{PBHs-Constraints}. 

\paragraph{Important Point:}So far we have considered a very wide range of the scales covered by PBH. Next, we figure out how much they do contribute on putting constraints on the amplitude of the primordial curvature power-spectrum. Surprisingly, as we will show in what follows, all of them do contribute almost the same and there is not any hierarchies between the constraints which are coming from any of the above current (or futuristic) observational constraints. However, there is indeed another type of the experiments, CMB spectral distortion, that gives us about three orders of magnitude stronger limits on the amplitude of the curvature perturbation, or even more for the proposed future PIXIE experiment. As the mechanism to derive the limit for this case is slightly different than the usual one for the above cases, we postpone considering this type of the constraints into the next sections and right after reading off the constraints on the amplitude of the power spectrum. Eventually we would compare them both to get a feeling on how they are gonna to put constraints on the primordial spectrum.

%%%%%%%%%%%%%%%%%%%%%%%%%%%%%%%%%%%%%%%%%%%%%%%%%%%

\begin{table}[ht]
	\caption{A summary of the constraints on the initial PBHs abundance, $\beta(M_{PBH})$ as a function of the mass. Possible future limits are designated in {\color{red}red}.}
	\label{tab:PBH}
%	\begin{tcolorbox}[tab2,tabularx={X||Y|Y||Y}]
%	\makebox[\linewidth][c]{\begin{tabular}	{|| c  c  c  c ||}
	\makebox[\linewidth][c]{\begin{tabular}	{@{}c  c c  c@{}}
			\hline 
			\hline
			{\small {\color{blue}{description}}} 
			& {\small {\color{blue}{wave number range}}}& \centering{{\small {\color{blue}{mass range}}}} & {\small{\color{blue}{constraints on $\beta(M_{PBH})$}}}\\
			\hline \\ 
			{\small {Disk heating}} 
			& {\small {$10^{-3}\lesssim (k/Mpc^{-1}) \lesssim 10^{3}$}}& \multirow{2}*{\small {$10^{7} \lesssim (M_{PBH}/M_{\odot}) \lesssim 10^{18}$}} & \multirow{2}*{\small  {$ \lesssim 10^{-3}\left(f_{M} \frac{M_{PBH}}{M_{\odot}}\right)^{-1/2}$}}\\[12pt]
			\hline \\
				{\small {Wide binary disruption}} 
				& {\small {$ 800 \lesssim (k/Mpc^{-1}) \lesssim 10^{5} $}}& {\small {$10^{3} \lesssim (M_{PBH}/M_{\odot}) \lesssim 10^{8} $}} & {\small  {$ \lesssim 6 \times 10^{-11}\left(\frac{M_{PBH}}{f_{M} M_{\odot}}\right)^{1/2}$}}\\[12pt]
				\hline \hline \\
					{{\color{red}\small {Fast Radio Bursts }}} 
					& {\small {$ 2.9 \times 10^{4} \lesssim (k/Mpc^{-1}) \lesssim 9.2 \times 10^{5} $}}& {\small {$ 10 \lesssim (M_{PBH}/M_{\odot}) \lesssim 10^{4} $}} & {\small  {$ \lesssim 1.4 \times 10^{-9} F_{D}(M_{PBH})\left(\frac{M_{PBH}}{M_{\odot}}\right)^{1/2}$}}\\[12pt]
				\hline \hline\\
				{\small {Quasar microlensing}} 
				& {\small {$ 1.2 \times 10^{5} \lesssim (k/Mpc^{-1}) \lesssim 6.5 \times 10^{7} $}}& {\small {$10^{-3} \lesssim (M_{PBH}/M_{\odot}) \lesssim 300 $}} & {\small  {$ \lesssim 2 \times 10^{-10}\left(\frac{M_{PBH}}{f_{M} M_{\odot}}\right)^{1/2}$}}\\[12pt]
				\hline \\
                {}
				& {\small {$ 4.5 \times 10^{5} \lesssim (k/Mpc^{-1}) \lesssim 1.42 \times 10^{6} $}}& {\small {$ 1 \lesssim (M_{PBH}/M_{\odot}) \lesssim 10 $}} & {\small  {$ \lesssim 6 \times 10^{-11}\left(\frac{M_{PBH}}{f_{M} M_{\odot}}\right)^{1/2}$}}\\[9pt]
			{\small {Microlensing}} & {\small {$ 1.42 \times 10^{6} \lesssim (k/Mpc^{-1}) \lesssim 1.4 \times 10^{9} $}}& {\small {$10^{-6} \lesssim (M_{PBH}/M_{\odot}) \lesssim 1.0 $}} & {\small  {$ \lesssim 2 \times 10^{-11}\left(\frac{M_{PBH}}{f_{M} M_{\odot}}\right)^{1/2}$}}\\[9pt]
				{} 	& {\small {$ 1.4 \times 10^{9} \lesssim (k/Mpc^{-1}) \lesssim 4.5 \times 10^{9} $}}& {\small {$ 10^{-7} \lesssim (M_{PBH}/M_{\odot}) \lesssim 10^{-6} $}} & {\small  {$ \lesssim  \times 10^{-10}\left(\frac{M_{PBH}}{f_{M} M_{\odot}}\right)^{1/2}$}}\\[9pt]
				\hline \\
					{\small {GRB femtolensing}} 
					& {\small {$ 4.5 \times 10^{12} \lesssim (k/Mpc^{-1}) \lesssim 1.4 \times 10^{14} $}}& {\small {$10^{-16} \lesssim (M_{PBH}/M_{\odot}) \lesssim 10^{-13} $}} & {\small  {$ \lesssim 2 \times 10^{-10}\left(\frac{M_{PBH}}{f_{M} M_{\odot}}\right)^{1/2}$}}\\[12pt]
					\hline \hline \\
					{\small {{\color{red}Reionization and $21 cm$ }}} 
					& {\small {$ 2 \times 10^{14} \lesssim (k/Mpc^{-1}) \lesssim 2 \times 10^{16} $}}& {\small {$10^{-20} \lesssim (M_{PBH}/M_{\odot}) \lesssim 10^{-16} $}} & {\small  {$ \lesssim 1.1 \times 10^{39}\left(\frac{M_{PBH}}{ M_{\odot}}\right)^{7/2}$}}\\[12pt]
					\hline \hline\\
					{\small {CMB SD (COBE/FIRAS)}} 
					& {\small {$ 2 \times 10^{16} \lesssim (k/Mpc^{-1}) \lesssim 2 \times 10^{17} $}}& {\small {$  10^{-22} \lesssim (M_{PBH}/M_{\odot}) \lesssim 10^{-20} $}} & {\small  {$ \lesssim 10^{-21}$}}\\[9pt]
					{{\color{red}\small {CMB SD (PIXIE)}}} & {} & {} & {\small  {$ \lesssim 10^{-24}$}}\\[12pt]
					\hline \\
					{\small {photodissociate D}} 
					& {\small {$ 2 \times 10^{16} \lesssim (k/Mpc^{-1}) \lesssim 6.3 \times 10^{17} $}}& {\small {$  10^{-24} \lesssim (M_{PBH}/M_{\odot}) \lesssim 10^{-21} $}} & {\small  {$ \lesssim 1.3 \times 10^{-10}\left(\frac{M_{PBH}}{f_{M} M_{\odot}}\right)^{1/2} $}}\\[12pt]
						\hline \\
					{\small {Hadron injection}} 
					& {\small {$ 6.3 \times 10^{17} \lesssim (k/Mpc^{-1}) \lesssim 6.3 \times 10^{18} $}}& {\small {$10^{-26} \lesssim (M_{PBH}/M_{\odot}) \lesssim 10^{-24} $}} & {\small  {$ \lesssim  10^{-20} $}}\\[12pt]
			        \hline \\
					{\small {Quasi-SMP}} 
					& {\small {$ 3.7 \times 10^{16} \lesssim (k/Mpc^{-1}) \lesssim 6.3 \times 10^{18} $}}& {\small {$10^{-26} \lesssim (M_{PBH}/M_{\odot}) \lesssim 10^{-21} $}} & {\small  {$ \lesssim 7 \times 10^{-30}\left(\frac{M_{PBH}}{f_{M} M_{\odot}}\right)^{-1/2} $}}\\[12pt]
					\hline \\
					{\small {LSP}} 
					& {\small {$ 2 \times 10^{17} \lesssim (k/Mpc^{-1}) \lesssim 2 \times 10^{21} $}}& {\small {$10^{-31} \lesssim (M_{PBH}/M_{\odot}) \lesssim 10^{-25} $}} & {\small  {$ \lesssim  7 \times 10^{-30}\left(\frac{M_{PBH}}{f_{M} M_{\odot}}\right)^{-1/2}  $}}\\[12pt]
					\hline \\
				    {\small {Planck relic}} 
				    & {\small {$ 2 \times 10^{21} \lesssim (k/Mpc^{-1}) \lesssim 7 \times 10^{25} $}}& {\small {$10^{-40} \lesssim (M_{PBH}/M_{\odot}) \lesssim 3 \times 10^{-31} $}} & {\small  {$ \lesssim 7 \times 10^{31}\left(\frac{M_{PBH}}{ M_{\odot}}\right)^{3/2} $}}\\[12pt]
				    \hline
				    \hline
		\end{tabular}}
%	\end{tcolorbox}
\end{table}
\clearpage
%%%%%%%%%%%%%%%%%%%%%%%%%%%%%%%%%%%%%%%%%%%%%%%%%%%
\subsection{Constraints on the primordial Power spectrum}
In the following, we focus on how to derive the constraints on the amplitude of the curvature perturbation using the presented upper limits on $\beta_{PBH}$. Throughout our subsequent calculations, we only consider the Gaussian initial condition. We leave the impact of different realizations of the Non-Gaussian initial conditions for a follow up work. \\
As we already discussed at the end of the above subsection, depending on the type of the observations, there are two different ways to put constraints on the amplitude of curvature perturbations. We consider both of these approach in what follows. The first way is suitable to put limit on the whole of different observational limits presented in Table. \ref{tab:PBH}. And, the second way is convenient for the Spectral Distortion type of the limits. 

\subsubsection{(a) The first way to put limit on curvature power spectrum}
\label{curvature-pow-PBH}
\begin{figure}[!h]
	\centering
	\includegraphics[width=0.96\textwidth]{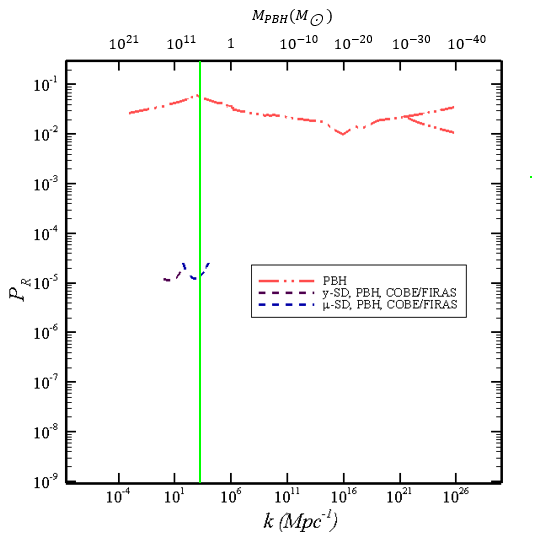}
	\caption{Different constraints on the amplitude of the primordial curvature perturbation power spectrum $P_R$ coming from PBHs versus wave number. 
		The solid green vertical line denotes %the largest PBH mass which is associated with the recombination, 
		$m_{PBH} = 10^5~M_{\odot}$.}
	\label{PBHs-Constraints}
\end{figure}

 Using this assumption, the probability distribution function for a smoothed density contrast would be given as,

\ba 
\label{PDF-density}
P\left(\delta_{hor}(R) \right) = \left(\frac{1}{\sqrt{2\pi}\sigma_{hor}(R)}\right) \exp{\left(-\frac{\delta^2_{hor}(R)}{2 \sigma^2_{hor}(R)}\right)}
\ea
here $\sigma_{hor}(R)$ refers to the mass variance as,

\ba
\label{mass-variance-PBH}
\sigma_{hor}^2(R) = \int_{0}^{\infty} \exp{\left(-k^2 R^2\right)} \mathcal{P}_{\delta}(k, t) \frac{dk}{k}
\ea
where we have used a Gaussian filter function as the window function.\\
Next, we use \textit{Press-Schechter} theory to calculate the initial abundance of PBH as,
\ba
\label{beta-initial-PBH}
\beta(M_{PBH}) &=& 2 f_{M} \int_{\delta_{crit}}^{1} P\left(\delta_{hor}(R) \right) d\delta_{hor}(R) \nonumber\\
&\sim & f_{M} erfc\left( \frac{\delta_{crit}}{\sqrt{2} \sigma_{hor}(R)}\right)
\ea
Eq. (\ref{beta-initial-PBH}) enables us to figure out the relationship between the initial abundance of the PBH and the mass-variance. \\
Finally, we use the following expression to translate this back to the initial curvature power-spectrum,
\ba
\label{power-delta}
\mathcal{P}_{\delta}(k, t) = \frac{16}{3} \left(\frac{k}{aH}\right)^2 j^2_{1}\left(\frac{k}{\sqrt{3}aH}\right) \mathcal{P}_{R}(k)
\ea
Plugging Eq. (\ref{power-delta}) back into Eq. (\ref{mass-variance-PBH}) and setting $R = aH$ we would get,

\ba
\label{mass-variance-PBH-2}
\sigma^2_{hor} (R) =  \frac{16}{3} \int_{0}^{\infty} \left(kR\right)^2 j^2_{1}\left(\frac{k R}{\sqrt{3}}\right) \mathcal{P}_{R}(k) \frac{dk}{k}
\ea
The above integral is dominated around $k \sim R^{-1}$ due to the presence of $j^2_{1}\left(\frac{k R}{\sqrt{3}}\right)$. In order to proceed, we would also assume that the power-spectrum can be locally approximated as a power-low function. In another word, we assume,
\ba
\label{initial-power-spectrum}
\mathcal{P}_{R}(k) = \mathcal{P}_{R}(k_{R}) \left(\frac{k}{k_{R}}\right)^{n(k_{R})-1} ~~~,~~~ k_{R} \equiv 1/R
\ea
We further assume that $n(k_{R})\sim 1$. We should emphasis here that relaxing this assumption does only change the results by few percent. So it is safe to neglect this dependency to simplify the analysis. The results of this analysis is given in Fig. \ref{PBHs-Constraints} red color.
%%%%%%%%%%%%%%%%%%%%%%%%%%%%%%%%%%%%%%%%%%%%%%%%%

%%%%%%%%%%%%%%%%%%%%%%%%%%%%%%%%%%%%%%%%%%%%%%%%%
\subsubsection{(b) The second way to put limit on curvature power spectrum}
\label{SD-PBH}
\begin{figure}[!h]
	\centering
	\includegraphics[width=0.96\textwidth]{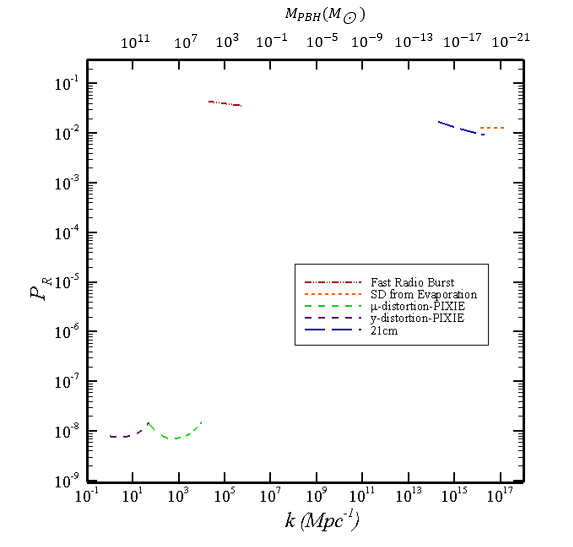}
	\caption{Futuristic constraints on the amplitude of the primordial curvature perturbation coming from PBHs.}
	\label{PBHs-Future}
\end{figure}

As we already mentioned above, there is another way to put (stronger) limits on the primordial power-spectrum. This mechanism is only suitable for some small scaled perturbations, within the window $ 1 \leqslant k/Mpc^{-1} \leqslant 10^{4}$, and arises from the Silk damping of the perturbations. This process generates $\mu$ and $y$ spectral distortions, \cite{Chluba:2012we}. Here we present, in some details, refers to \cite{Chluba:2012we} for more detailed analysis, how to use these distortions to put constraints on the amplitude of the curvature power-spectrum. For this purpose, we would first present an expression for the $\mu$ and $y$ distortions, 
\ba
\label{mu-dis}
\mu &=& 13.16 \int_{z_{\mu,y}}^{\infty} \frac{dz}{(1+z)} \mathcal{J}_{bb}(z) \int \frac{k dk}{k^2_{D}} \mathcal{P}_{R}(k_{R}) \exp{\left(-2\left(\frac{k^2}{k^2_{D}(z)}\right)\right)} W(k/k_{R}) \\
\label{y-dis}
y &=& 2.35 \int_{0}^{z_{\mu,y}} \frac{dz}{(1+z)} \int \frac{k dk}{k^2_{D}} \mathcal{P}_{R}(k_{R}) \exp{\left(-2\left(\frac{k^2}{k^2_{D}(z)}\right)\right)} W(k/k_{R})
\ea
here $k_{D} = 4.1 \times 10^{-6} (1 + z)^{3/2} Mpc^{-1}$ is the silk damping scale. In addition, $\mathcal{J}_{bb}(z) = \exp{\left(- \left(z/z_{\mu}\right)^{5/2}\right)}$ denotes the visibility function for the spectral distortion where $z_{\mu} = 1.98 \times 10^{6}$ and $z_{\mu,y} = 5 \times 10^{4}$ is the transition redshift from $\mu$ distortion to $y$ distortion. Furthermore, $W(k/k_{R})$ refers to the window function. We use Gaussian and top hat filter functions for PBH as well as the UCMHs, that we consider in the following sections, respectively. \\ 
The upper limits on the $\mu$ and $y$ distortions coming from the COBE/FIRAS experiments are
\ba
\label{COBE-FIRAS}
\mu &\leqslant&  9 \times 10^{-5} ~~~~~~~~,~~~~~~~~~
y \leqslant  1.5 \times 10^{-5} 
\ea
Using Eqs. (\ref{mu-dis}-\ref{COBE-FIRAS}), we can then calculate the upper limits on the curvature power spectrum. 
The results are given in Fig. \ref{PBHs-Constraints}, dashed purple and cyan colors. \\
From the plot it is clear that there are at least three orders of magnitude hierarchy between the constraints coming from the spectral distortion as compared with the rest of the experiments, i.e. those who have been discussed in subsection \ref{PBH-Experiments}. Therefore for PBHs CMB spectral distortion could be more informative. In addition, we should emphasis here that we only considered COBE/FIRAS type of the experiments, Eq. (\ref{COBE-FIRAS}). Going beyond that and considering PIXIE-like experiments could push this upper limit by another three orders of magnitude down, \cite{Chluba:2012we}, and potentially bring a new window for the early Universe cosmology.\\
 Before closing this section, we also present the futuristic constraints to see how much improvements we could expect going into futuristic experiments. According to Table \ref{tab:PBH} there are three different ongoing experiments. In Fig. \ref{PBHs-Future} we present their constraints on the curvature power-spectrum. In addition, we also present the possible improvements in the constraints from the PIXIE-like experiment. As we see the PIXIE could push the constraints by three orders of magnitude. 

%%%%%%%%%%%%%%%%%%%%%%%%%%%%%%%%%%%%%%%%%%%%%%%%%

%%%%%%%%%%%%%%%%%%%%%%%%%%%%%%%%%%%%%%%%%%%%%%%%%

\section{UCMHs and inferred limits on the primordial curvature power spectrum}
\label{UCMH-introduction}
By definition, UCMHs are dense dark matter structures, which can be formed from the larger density perturbations right after the matter-radiation equality. 
As was suggested in \cite{Ricotti:2009bs}, if DM inside this structure is in the form of Weakly Interacting Massive Particles, WIMPs, UCMHs could be detected through their Gamma-Ray emission. Scott and Sivertsson, \cite{Scott:2009tu}, 
the $\gamma$-emission from UCMHs could happen at different stages, such as $e^{-}e^{+}$-annihilation, QCD and EW phase-transitions. This means that there should be an upper-limit cut-off on the available scale that we consider. 
On the other word, this kind of observations can not be as sensitive as that of PBH where we could go to the comoving wave-number as large as $10^{20} Mpc^{-1}$. On the other hand, considering WIMPs as DM particles, there is also another (stronger) limit on the small scales structures that can be probed this way. 
As was first pointed out in \cite{Bringmann:2009vf}, the kinetic decoupling of WIMPs in the early universe would put a small-scale cut-off in the spectrum of density fluctuations. 
Assuming a decoupling temperature of order $MeV$-$GeV$, the smallest protohalos that could be formed would range up between $10^{-11}-10^{-3} M_{\odot}$. So in average the smallest possible mass would be about $10^{-7} M_{\odot}$. Translating this into the scale, we would get an average upper limit on the wave number of order $10^{7} Mpc^{-1}$. Interestingly, this corresponds to a time between the QCD and EW phase-transition which is somehow consistent with our previous thought about the upper-limit on the wave-number. It is worth to emphasis here that such an upper limit could be a unique way of shedding light about the nature of the DM  as well as the fundamental high-energy universe. So it would be interesting to propose some experiments that are very sensitive to this cuf-off. \\
%{\color{red} Add few more sentences about the kinetic decoupling later!}

As they pointed out, density perturbations of the order $10^{-3}$, though its exact number is scale dependent as we will point it out in what follows, can collapse prior or right after the matter-radiation equality and therefore seed the formation of UCMHs. \\
Before going through the details of the formation as well as the structure of UCMHs, we compare them with both of the normal fluctuations during inflation as well as PBHs. 
As it is well-known, for most of the inflationary models, the density fluctuations are so tiny, about $10^{-5}$, at the Horizon entry. So they do not collapse until some time after the matter-radiation equality. 
However, there are some inflationary models in which the power-spectrum is much larger in some particular scales. 
Therefore the structures of the size associated with this particular scale might collapse far earlier than the usual case, even before the matter-radiation equality. 
The most (known) example of such a very rapid collapse occurs for PBHs. 
In fact, PBHs would form once a perturbation of required amplitude get back into the horizon. 
A less severe case is UCMHs, in which the density perturbation gets back into the horizon during the radiation dominance and  it  would collapse at $z > 1000$ at least after matter-radiation equality.  \\
\subsection{Formation and the structure of UCMHs}
Density fluctuations about $10^{-3}$ during Radiation Dominance suffice to create over-dense regions which later would collapse to UCMHs provided they survive to Matter dominance era. The corresponding mass for this specific fluctuations at the horizon re-entry would be, 
\ba
\label{initial-mass-UCMH}
M_{i} &\simeq& \left(\frac{4\pi}{3} \rho_{\chi} H^{-3}\right)\Bigg{|}_{aH \sim \frac{1}{R}} \nonumber\\
&=& 1.3 \times 10^{11} \left(\frac{\Omega_{\chi}h^2}{0.112}\right) \left(\frac{R}{Mpc}\right)^3 M_{\odot}
\ea
here $\Omega_{\chi}$ denotes the fraction of the critical over-density in the form of the DM today. \\
As in the case of PBHs,  identify $R$ with $1/k$. Doing this we would get,
\ba
\label{Initial-mass-k}
M_{i} \simeq 1.3 \times 10^{11} \left(\frac{\Omega_{\chi}h^2}{0.112}\right) \left(\frac{{Mpc}^{-1}}{k}\right)^{3} M_{\odot}
\ea
Comparing Eq. (\ref{Initial-mass-k}) with Eq. (\ref{mass-range}), we see that their scale dependent is different. This is because, for PBH we use the radiation energy-density while for UCMHs we use the matter energy density. \\
During RD, the above mass is almost constant. However, it starts growing from the matter-radiation equality, both due to the infall of the matter as well as the baryons, as
\ba 
\label{mass-increase}
M_{UCMH}(z) = \left(\frac{1+z_{eq}}{1+z}\right) M_{i} 
\ea
A conservative assumption is that such a growth would continue until the epoch of standard structure formation which happens around $z \sim 10$. Therefore, the current mass of UCMH is given as,
\ba
\label{current-mass-UCMH}
M^{0}_{UCMH} &\equiv& M_{UCMH}(z \leqslant 10) \nonumber\\
&\approx& 4 \times 10^{13} \left(\frac{R}{Mpc}\right)^3 M_{\odot}
\ea
where we have assumed $\Omega_{\chi}h^2 = 0.112$. \\
Finally, for our next references, it is interesting to express% the relationship between $f_{UCMH} \equiv \frac{\Omega_{UCMh}}{\Omega_{DM}}$ as well as the initial abundance of the UCMH, i.e. $\beta(M_{H}(z_{i}))$,
\ba
\label{fraction-abundance-UCMH}
f_{UCMH} &\equiv& \frac{\Omega_{UCMh}(M^{(0)}_{UCMH})}{\Omega_{DM}} = \bigg{[}\frac{M_{UCMH}(z =0)}{M_{UCMH}(z_{eq}}\bigg{]}~\beta(M_{H}(z_{i})) \nonumber\\
&=&\left(\frac{400}{1.3}\right)\beta(M_{H}(z_{i}))
\ea

\subsection{Constraining UCMH abundance from different Astro-particle-physical experiments}
Having presented a quick introduction to UCMHs, it is worth figuring out how to observe or limit them and several different ways to (uniquely) distinguish them from the rest of the DM candidates. \\
In what follows, we present few different possibilities to shed light about UCMHs. As we will see, a null result out of each of these tests will lead to an upper-limit on the fraction of the DM in the form of the UCMHs. Finally, we would try to connect these upper limits to an upper-bound on the primordial curvature power-spectrum.  \\
Before going through the details of these tests though, let us divide them up in two different categories, 
\textbf{\textit{$(1)$ Observational signals sensitive to the nature of the DM particles within UCMH structures}}\\
These are several items that are belong to this category as,
\begin{itemize}
\item (Non-)observation of Gamma- rays from UCMHs by the Fermi-LAT, 
\item (Possible) Neutrino signals from UCMHs to be detected by IceCube/DeepCore,
\item (Any) Modifications of Reionization observed by WMAP/Planck.
\end{itemize}
\textbf{\textit{$(2)$ Observational signals which are NOT sensitive to the nature of the DM particles within UCMH structures}}\\
There are also (few) different mechanisms belong to this group as, 
\begin{itemize}
\item (Potential) Lensing time delay in Pulsar Timing from UCMHs to be detected in ATNF pulsar catalogue,
\item (Upper-limits) on CMB Spectral Distortion from UCMH by COBE/FIRAS and future observations.
\end{itemize}
%%%%%%%%%%%%%%%%%%%%%%%%%%%%%%%%%%%%%%%%%%%%%%%%

\begin{figure}[!h]
	\centering
	\includegraphics[width=0.8\textwidth]{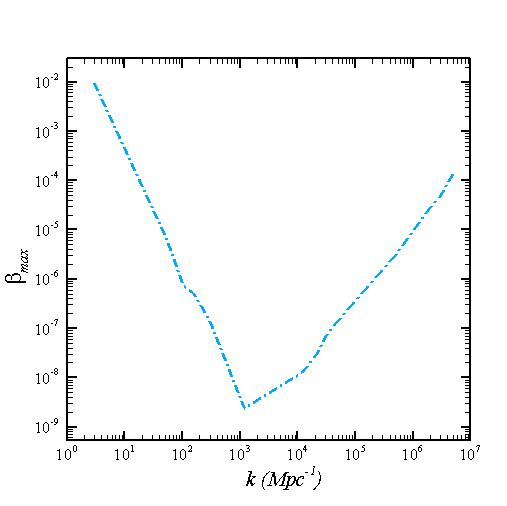}
	\caption{The upper bound on the abundance of UCMHs from Fermi-LAT.}
	\label{abundance-UCMH-Fermi}
\end{figure}
%%%%%%%%%%%%%%%%%%%%%%%%%%%%%%%%%%%%%%%%%%%%%%%%%
Below we would consider the above ansatzs in some details and present the upper bounds on the abundance of UCMHs from each of them. 

\subsubsection{ Gama-ray searches from UCMHs by Fermi-LAT}
As we already mentioned, one of the most promising features of UCMHs was the fact that they could be a source of the Gamma rays if the DM inside of them is made of WIMPs. More specifically, $\gamma$-rays could be produced out of WIMP annihilation either to $\mu^{-}\mu^{+}$ or to $b\bar{b}$. So in principle, their contribution must be added together to give us the Flux of the photons. However, in order to find the most severe constraints on the abundance of UCMHs from the Non-detection of $\gamma$-rays, in \cite{Carr:2009jm} the authors only considered the annihilation of WIMP to $b\bar{b}$. The reason is that the photon flux for $\mu^{-}\mu^{+}$ is smaller than that for $b\bar{b}$. Therefore, the maximum available value for the abundance of UCMHs without big enough $\gamma$-rays is smaller for $b\bar{b}$ as compared with $\mu^{-}\mu^{+}$. In fact, it goes like $f_{max} \propto \Phi^{-3/2}$, see \cite{Carr:2009jm} for more details. Intuitively, it does make sense. Because the more the flux is the more constraint would be the abundance for no-detection experiments. Having this said, in what follows, we also only present the constraints for the $b\bar{b}$ annihilation. In addition, we only consider WIMP with mass $m_{\chi} = 1 TeV$ with the cross-section $\langle \sigma v\rangle = 3 \times 10^{-26} cm^{3} s^{-1}$.  Observationally, since there has not been any detection of the $\gamma$-rays out the DM decay by Fermi, \cite{Buckley:2010vg, Ackermann:2012nb}, we could find an upper limit on the initial abundance of UCMHs within the Milky Way, \cite{Klypin:2001xu, Josan:2010vn}. The results is shown in Fig. \ref{abundance-UCMH-Fermi}.
%%%%%%%%%%%%%%%%%%%%%%%%%%%%%%%%%%%%%%%%%%%%%%%%%
\begin{figure}[!h]
	\centering
	\includegraphics[width=0.8\textwidth]{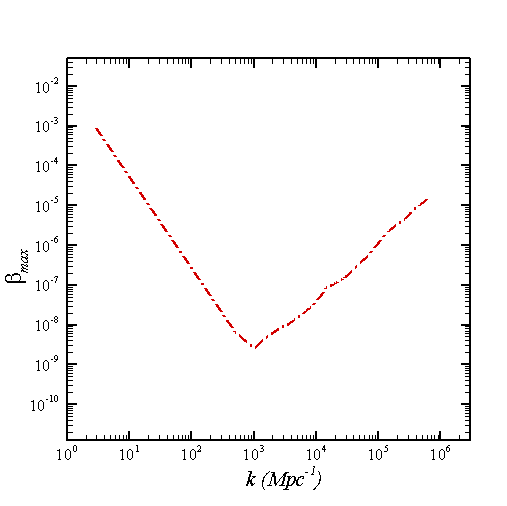}
	\caption{The upper bound on the abundance of UCMHs from IceCube/DeepCore.}
	\label{abundance-UCMH-Neutrino}
\end{figure}
%%%%%%%%%%%%%%%%%%%%%%%%%%%%%%%%%%%%%%%%%%%%%%%%%
\subsubsection{Neutrino Signals from UCMHs by IceCube/DeepCore}
If DM inside the UCMHs are made of WIMPs, in addition to $\gamma$-rays we should also get Neutrinos; being produced from the WIMP annihilation. These are expected to be detected either by ICeCube or DeepCore \cite{Yang:2013dsa}. Moreover, the production of the $\gamma$-rays is accompanied with Neutrinos when the dark matter annihilate. Therefore, the current search for the Neutrinos are indeed  complementary to that of the photons. 
Though we have three types of Neutrinos, among them $\nu_{\mu}$ is the target for this search. 
The reason is that propagating through the matter or getting into the detectors, $\nu_{\mu}$ could be converted to muons ( $\mu$). These are called the upward and contained events, respectively. Later, muons could be observed by the detectors on the Earth through their Cherenkov radiation. Here we only consider those muons which are produced inside the detectors, contained events. However, it turns out that the upward events also lead to nearly the same constraints, \cite{Yang:2013dsa}. In addition, we only consider WIMP with the mass $m_{\chi} = 1 TeV$ and with the averaged cross-section $\langle \sigma v\rangle = 3 \times 10^{-26} cm^{3} s^{-1}$. The upper limit on the initial abundance of UCMH is shown in Fig. \ref{abundance-UCMH-Neutrino}.

\subsubsection{Change in Reionization Observed by WMAP/Planck }
\begin{figure}[!h]
	\centering
	\includegraphics[width=0.78\textwidth]{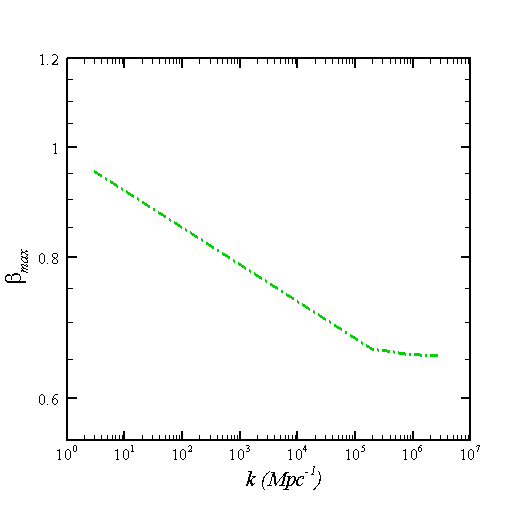}
	\caption{The upper bound on the abundance of UCMHs from the Reionization coming from CMB observations.}
	\label{abundance-UCMH-Reionization}
\end{figure}

Another interesting fingerprint of WIMP annihilation is their impact on Reionization Epoch as well as the integrated optical depth of the CMB. If DM is within UCMHs, their impact on the evolution of Intergalactic Medium is more important. This structures could ionize and heat up the IGM after the matter-radiation equality which is much earlier than the formation of the first stars, \cite{Zhang:2010cj}. This effect happens around the equality. So in connecting this into the fraction of the DM in UCMHs now we have $f_{UCMH}(z =0) \simeq 340 f_{UCMH}(z_{eq})$. As for $f_{UCMH}(z_{eq})$ we have, 
\ba
\label{f-equality}
f_{UCMH}(z_{eq}) \lesssim 10^{-2} \left(\frac{m_{\chi}}{100~GeV}\right)
\ea
This means that for $m_{\chi} = 1 TeV$, $f_{UCMH}$ would be saturated before today. 
From this and Using Eq. (\ref{fraction-abundance-UCMH}), we could then achieve an upper limit on the initial abundance of UCMH. 
This upper bound is presented in Fig. \ref{abundance-UCMH-Reionization}. 
The presented plot is coming from the WMAP results. 
Adding the recent results from the Planck, could push the upper limits on the initial abundance by at most two orders of the magnitude, \cite{Chen:2016pud}. However, things does not change too much as the limits coming from the Reionization are indeed the most relaxed ones as compare with the rest of the limits.
\\
Having considered the details of the tests which are sensitive to the nature of the DM, in what follows, we would focus on few more events which are blind to the type of the DM within the UCMHs. While the first one is only sensitive to the gravity, the pulsar timing delay, the second one described by any distortion in the CMB SD. 

%%%%%%%%%%%%%%%%%%%%%%%%%%%%%%%%%%%%%%%%%%%%%%%%%

%%%%%%%%%%%%%%%%%%%%%%%%%%%%%%%%%%%%%%%%%%%%%%%%%
\subsubsection{Lensing time delay in Pulsar Timing by ATNF catalogue}
\begin{figure}[!h]
	\centering
	\includegraphics[width=0.78\textwidth]{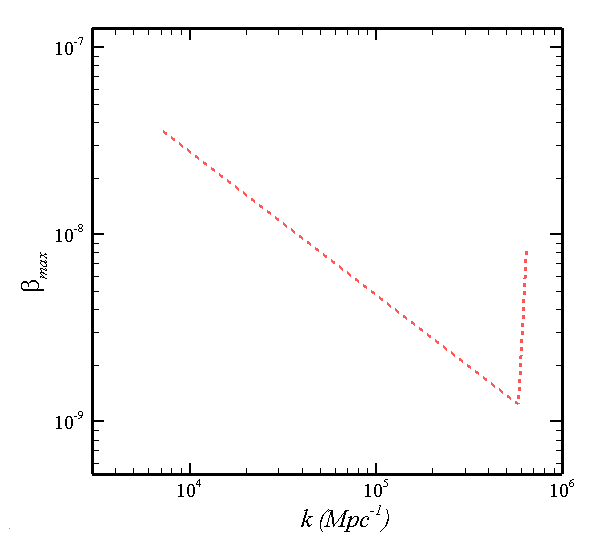}
	\caption{The upper bound on the abundance of UCMHs from the Lensing time delay in Pulsar Timing by ATNF catalog.}
	\label{abundance-UCMH-Pulsar}
\end{figure}

There are some tests based on the gravitational effects and therefore do not care about the nature of the DM. 
There are at least two different types of the experiments like this. 
The first one is the
\textit{Gravitational Lensing} known as a very important tool to detect DM. 
And, it can put constraints on the abundance of the UCMHs by non-observing any changes in the position and/or the light curve of stars. However, as it turns out, in order to increase the sensitivity of such a test, one does need to consider super-precise satellites like THEIA. Another, even more precise, way to shed light about UCMHs is using the time delay in \textit{Pulsar Timing}. In another word, we would like to measure the effect of an intermediate mass, here UCMHs, on the arrival time of millisecond pulsar, \cite{Clark:2015sha}.
This method is based on the Shapiro effect, i.e. measuring any changes in the travel time of light rays which are passing through an area in which the gravitational potential is changing.  
As it was pointed out in \cite{Clark:2015sha}, practically, one should take a look at the frequency of a pulse over a period of time, say several years. And wait for any decrease, when a DM halo moves toward the line of sight of the pulse, as well as a subsequent increase, happens when it moves away from that area. 
Since millisecond pulsars are very accurate clocks, they can measure very small effects as well. 
Although this effect is tiny for an individual masses, it is indeed sizeable for a population of DM halo objects. 
More precisely, such a structure would lead to an increase in the dispersion of the measured  \textit{period derivative} of the pulsar. Using the statistical results of ATNF pulsar catalog, one finds an upper bound on the value of this dispersion 
and so forth on the abundance of the UCMHs. This bound can be translated back onto an upper bound on the number density of UCMHs within Milky Way and therefore put constraints on the initial abundance of UCMH. The results are shown in Fig. \ref{abundance-UCMH-Pulsar}.
\footnote{There was an Erratum on the first publication of this bound. We thank Simon Bird for pointing this out to us.}

%%%%%%%%%%%%%%%%%%%%%%%%%%%%%%%%%%%%%%%%%%%%%%%%%

%%%%%%%%%%%%%%%%%%%%%%%%%%%%%%%%%%%%%%%%%%%%%%%%%

\subsubsection{Limits on CMB Spectral Distortion by COBE/FIRAS}
\begin{figure}[!h]
	\centering
	\includegraphics[width=0.78\textwidth]{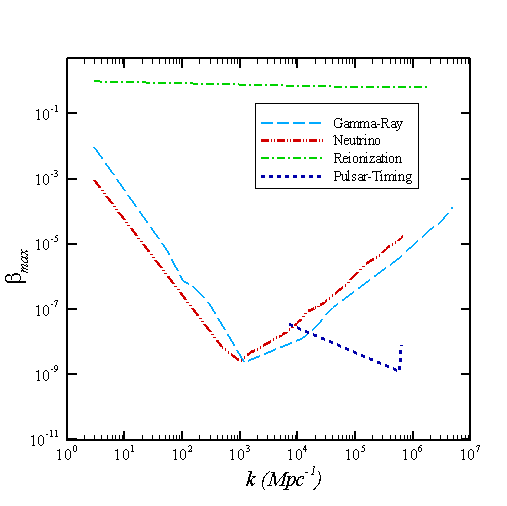}
	\caption{A comparison between the upper bound on the abundance of UCMHs from different tests, such as Gamma-Ray, Dashed blue curve, Pulsar-Timing, Dotted orange curve, Neutrinos, dashed dot-dot red curve, and from Reionization, dashed dotted green curve.}
	\label{abundance-UCMH-Combined} 
\end{figure}

As we already discussed above, in section \ref{SD-PBH}, CMB spectral distortion could be also used as a way to shed-light about the DM. This is also independent of the nature of the DM. Indeed the analysis is quite similar to the case of PBHs, with only replacing the Gaussian filter function with the top-hat window function. So we skip repeating this again and just present the limits on the power-spectrum in the following part. The bounds coming from the CMB SD on the UCMHs are of the same order as in the case of the PBHs. Again, COBE/FIRAS give us more relaxed constraints while the limits coming from the proposed PIXIE would be more severe by about three orders of the magnitude.\\
Before going through the details of the constraints on the power-spectrum, it is worth comparing the upper-bounds on the initial abundance of UCMHs for different tests. The results of this comparison is shown in Fig. \ref{abundance-UCMH-Combined}.

%%%%%%%%%%%%%%%%%%%%%%%%%%%%%%%%%%%%%%%%%%%%%%%%%
%%%%%%%%%%%%%%%%%%%%%%%%%%%%%%%%%%%%%%%%%%%%%%%%%

\subsection{Constraints on the primordial Power spectrum}
Having presented different constraints on the initial value of abundance, in what follows, we would translate them into an upper limit on the amplitude of the initial curvature perturbation. Likewise the case of PBHs, which was considered in subsection \ref{curvature-pow-PBH}, we assume Gaussian initial condition. So the probability distribution function would be similar to Eq. (\ref{PDF-density}). However, in this case the variance would not be the same as Eq. (\ref{mass-variance-PBH}). Instead of that, it would behave as, \cite{Carr:2009jm},
\ba
\label{mass-variance-UCMH}
\sigma^2_{hor}(R) = \int_{0}^{\infty} W^2_{TH}(kR) \mathcal{P}_{\delta}(k) \frac{dk}{k}
\ea
where $W_{TH}$ denotes the top-hat window function which is given as,
\ba
W_{TH}(x) \equiv 3 \left(\frac{\left(\sin{x} - x\cos{x}\right)}{x^3}\right)
\ea
In addition, $\mathcal{P}_{\delta}(k)$ refers to the power-spectrum of the density fluctuations.
During the radiation dominance the matter density perturbations do have the following evolution, \cite{Carr:2009jm},
\ba
\label{density-evolv}
\delta(k, t) =  \theta^2 T(\theta) \mathcal{R}^{0}(k)
\ea
where $\theta = \frac{1}{\sqrt{3}} \left(\frac{k}{aH}\right)$. Moreover, $T(x)$ denotes the transfer function which turns out to have the following expression,
\ba
T(\theta) = \frac{6}{\theta^2} \bigg{[} \ln{(\theta)} + \gamma_{E} - \frac{1}{2} - Ci(\theta) + \frac{1}{2} j_{0}(\theta) \bigg{]}
\ea
%%%%%%%%%%%%%%%%%%%%%%%%%%%%%%%%%%%%%%%%%%%%%%%%%
\begin{figure}[!h]
	\centering
	\includegraphics[width=0.78\textwidth]{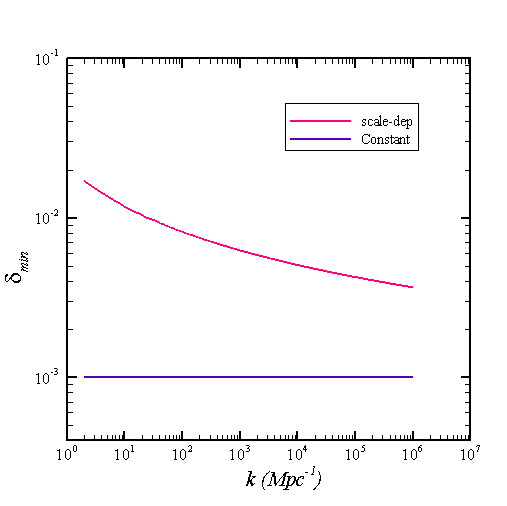}
	\caption{A comparison between two different choices for the $\delta_{min}$. }
	\label{delta-mins} 
\end{figure}
%%%%%%%%%%%%%%%%%%%%%%%%%%%%%%%%%%%%%%%%%%%%%%%%%
here $\gamma_{E} = 0.577216$ is the Euler-Mascheroni constant, $Ci$ is the Cosine integral and finally $j_{0}$ denotes the spherical Bessel function of the first rank. \\
Our next task is calculating the initial abundance for UCMHs. It is similar to Eq. (\ref{beta-initial-PBH}) for PBHs with changing the minimal required value of density contrast, i.e. $\delta_{min}$. As we pointed it out before, for producing PBH $\delta_{min} \simeq 1/3$. However, for in UCMH case things are slightly more complicated and there are a bit ambiguities in the literature. While some of the authors approximate it to be around $0.001$, the others also take into account the scale and the red-shift dependent of this function and achieve the following expression for $\delta_{min}$ as, 
\ba
\label{delta-min-UCMHs-sz}
\delta_{min}(k, t) = \frac{8}{9} \left(\frac{3\pi}{2}\right)^{2/3} \left(a^2H^2\bigg{|}_{z=z_c}\right) \frac{T(1/\sqrt{3})}{k^2 \mathcal{T}(k)} 
\ea
where $z_{c} $ denotes the red-shift of the collapse fr UCMHs. In addition, $\mathcal{T}(k)$ refers to a fitting formula for the transfer function around the time of matter-radiation equality, with the following form, 
\ba
\label{transfer-function}
\mathcal{T}(\kappa) &\simeq& \left( \frac{\ln{\left(1 + (0.124 \kappa)^2\right)}}{(0.124 \kappa)^2} \right) \nonumber\\
&\times& \bigg{[}
\frac{1 + (1.257 \kappa)^2 + (0.4452 \kappa)^4 + (0.2197 \kappa)^6}{1 + (1.606 \kappa)^2 + (0.8568 \kappa)^4 + (0.3927 \kappa)^6}
 \bigg{]}^{1/2}
\ea
where $\kappa \equiv \left(\frac{k\sqrt{\Omega_{r}}}{H_0 \Omega_m}\right)$. \\
It is now worth to do compare the above two choices for $\delta_{min}$ with each other. We have presented this in Fig. \ref{delta-mins}.

%%%%%%%%%%%%%%%%%%%%%%%%%%%%%%%%%%%%%%%%%%%%%%%%%
\begin{figure}[!h]
	\centering
	\includegraphics[width=0.93\textwidth]{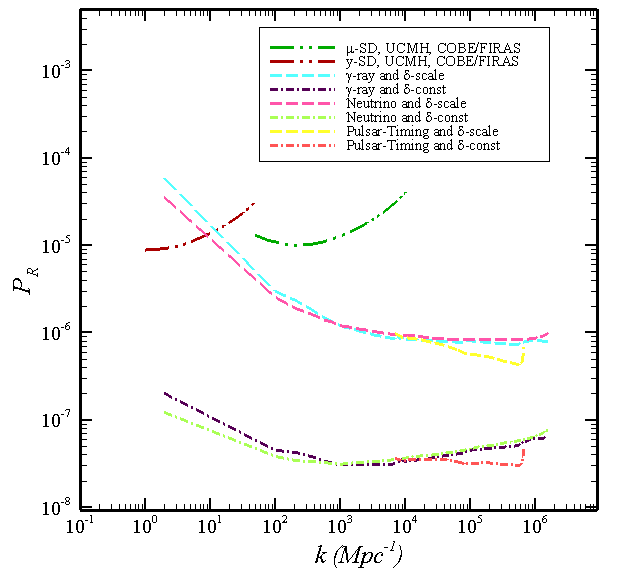}
	\caption{The upper-limit on the amplitude of the curvature power spectrum from UCMHs. Here $\delta$-scale means the scale dependent $\delta_{min}$ while $\delta$-const means that we have used $\delta_{min} = 0.001$. The dash-dot-dot lines denote the constrains coming from the current limits from COBE/FIRAS.}
	\label{power-ucmh} 
\end{figure}
%%%%%%%%%%%%%%%%%%%%%%%%%%%%%%%%%%%%%%%%%%%%%%%%%
Taking the above points into account, the initial abundance of UCMHs would be,
%%%%%%%%%%%%%%%%%%%%%%%%%%%%%%%%%%%%%%%%%%%%%%%%%
\begin{figure}[!h]
	\centering
	\includegraphics[width=0.93\textwidth]{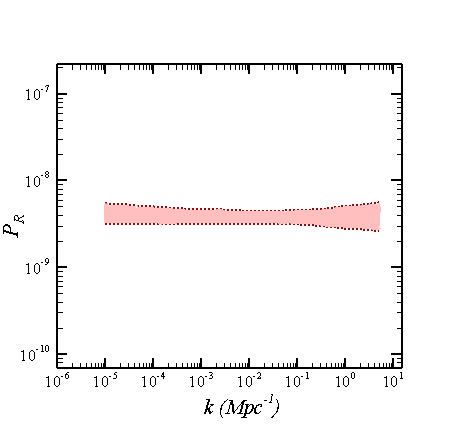}
	\caption{The band of allowed  curvature power spectrum amplitude from the combination of CMB, LSS and Ly$\alpha$.  }
	\label{power-cmb-lss-lya} 
\end{figure}
%%%%%%%%%%%%%%%%%%%%%%%%%%%%%%%%%%%%%%%%%%%%%%%%%
\newpage
\ba
\label{abundance-UCMH}
\beta(R) &\simeq& \left(\frac{\sigma_{hor}(R)}{\sqrt{2\pi} \delta_{min}} \right)\exp{\left(-\frac{\delta^2_{min}}{2 \sigma^2_{hor}(R)}\right)} \nonumber\\
&\sim& erfc\left(\frac{\delta_{min}}{\sqrt{2}\sigma_{hor}(R) }\right)
\ea

Inversing Eq. (\ref{abundance-UCMH}) and using the above upper-limits on the initial abundance of UCMHs, we could calculate the upper-limits on the amplitude of the curvature perturbations for each of the above tests. \\
The results are given in Fig. \ref{power-ucmh}. There are several important points which are worth to be mentioned, \\
\begin{itemize}
\item There is a hierarchy of about three orders of magnitude between the constraints which are coming from the scale dependent $\delta_{min}$ as compared with the constant $\delta_{min}$. Indeed the second choice gives us more constrained results for the curvature power-spectrum.	
\item Quite interestingly, the constraints from the Gamma-ray, Neutrinos and Pulsar-Timing are of the same order. Both for the constant and the scale dependent $\delta_{min}$. 
\item The constrained from Reionization are more relaxed as compared with the rest of the tests.
\item The current version of the CMB spectral distortion, coming from COBE/FIRAS, does give us comparable results as compared with the scale-dependent $\delta_{min}$. As the next generation of CMB spectral distortion, from PIXIE, is supposed to push down the sensitivity by (about) three orders of magnitude, we do expect that their results are 
comparable with the constraints coming from constant value of $\delta_{min}$. 
Therefore, even if we do not assume a constant value for $\delta_{min} = 0.001$, by going to the next generation of CMB experiments, we could hopefully get to a level of about $10^{-7}$ constraints on the amplitude of the curvature perturbation. 
\end{itemize}

\section{Determining the Curvature perturbation from CMB, Lyman-$\alpha$ and LSS}
\label{cmb-lss-lya-constraints}
Our final consideration would be on the most severe constraints on the relatively large scales probed by inflation. 
Here we cover the scales in the range $10^{-5} Mpc^{-1} \le k \le 1 Mpc^{-1}$. 
The CMB, large-scale structure and Ly$\alpha$ observations provide the best detections. 
We just present the final results here. The details can be found in \cite{Ravenni:2016vjd, Armengaud:2017nkf}. \\
The results of the constraints are given in Fig. \ref{power-cmb-lss-lya}.\\
\section{Constraining the Dark Matter mass fraction for PBHs and UCMHs}
So far we have presented the constraints on the initial abundance as well as the primordial power-spectrum for both PBHs and UCMHs.
We now translate the above limits into the constraints on the fraction of the DM, 
$f_{DM}$, to be made of either PBH or the UCMHs at different mass scales. \\
It is also very interesting to challenge ourself to figure out how, and under which conditions, the tight constraints on the amplitude of the curvature perturbations from the UCMHs sector lead to "additional NEW constraints" 
on the fraction of the DM in PBH form. 
More explicitly, we would like to see whether the limits on UCMHs can also be treated to a limit on the PBH $f_{DM}$.\\
In the following subsections, we consider these two questions separately. 
\subsection{Direct constraints on $f_{DM}$ in both of PBHs and the UCMHs}
\begin{figure}[!h]
	\centering
	\includegraphics[width=1.1\textwidth]{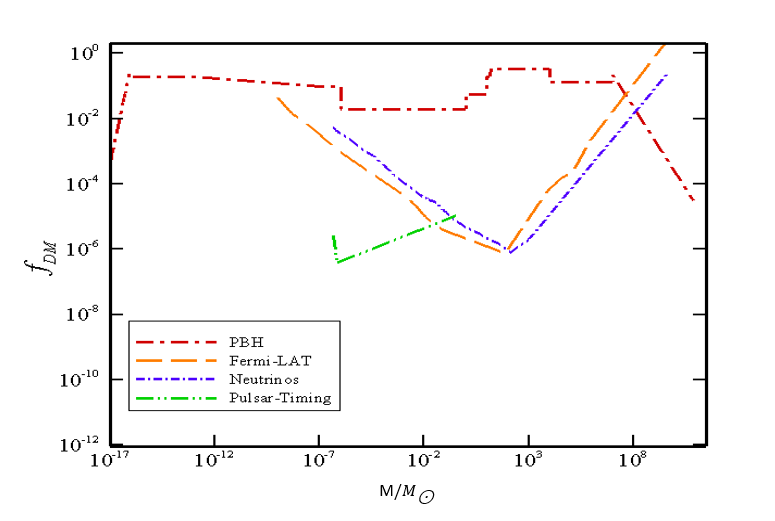}
	\caption{The upper-limit on the amplitude on the fraction of the dark matter contained in PBHs and UCMHs. 
	The pulsar timing limits apply to PBHs as well as UCMHs.}
	\label{combinedfdm} 
\end{figure}

Let us begin with the direct constraints on the fraction of the DM for both of the PBH and the UCMHs. This can be done by trying to translate the limits on the initial abundance into  $f_{DM}$. More explicitly, for the case of PBH, $f_{max}$ is found by using Table. \ref{tab:PBH}, for the information about the $\beta$ and then by using Eq. (\ref{beta-fh}) to calculate  $f_{max}$. 
For the UCMHs, we use the combination of Fig. \ref{abundance-UCMH-Combined} as well as the revised version of Eq. (\ref{fraction-abundance-UCMH}) to calculate the value of  $f_{max}$.
We present the limits of the $f_{DM}$ in Fig. \ref{combinedfdm}. \\
%%%%% Assuming that remake the figure with same mass scale I comment these statements out - 26 May GFS
% As the mass scaling of PBHs differs from that for UCMHs, for a given wave number $k$, we have two axes to denote the mass of the object as aligned by wave number $k$. \\
%While the lower mass axes refers to the mass of the UCMHs, the top part axes denotes the mass of the PBHs. 
From the plot, it is clear that the constraints on the UCMHs are more severe  than the PBHs by several orders of the magnitude. \\
\subsection{Indirect constraints on $f_{DM}$ of PBH due to tight limits on the UCMHs}	
Having presented the direct limits on the fraction of the DM in the form of the compact object, 
we now determine how much more information can we get from the limits on UCMHs in regard to limits on PBHs. 
How do the tight constraints on the $f_{DM}$ and the primordial power spectrum for UCMHs affect the bounds on the PBHs. One should be careful that such a link depends on the assumption that both objects have the same production mechanism. 
In other words, they must have been originated from the same seeds. 
For example they should both come from the primordial curvature fluctuations 
which is one of the most natural mechanisms to create them up. 
Assuming this happens, we will not expect too much hierarchies between these two. 
This is directly related to the tighter constraints on the primordial curvature power spectrum. 
We can conclude that if they came from the same primordial perturbations, then the failure to obtain UCMHs
which require perturbations at $\leq 10^{-3}$ level, would imply that any reasonable PDF, not only Gaussian,
would predict very low expectation of perturbations at 0.3 level.
That is to say if the PBHs come from Inflationary/primordial fluctuations, 
then one would generally anticipate that there would be many more UCMHs than corresponding PBHs
and their mass contribution to $f_{DM}$ is much less,
Thus the curvature fluctuation limit and the $f_{DM}$ would restrict the fraction of DM in PBH between about  
$1 M_\odot  \leq M_{PBH} \leq 10^{5.5} M_\odot$ to be much lower than $f_{DM} \leq 10^{-2}$.\\
In order to have an intuitive picture on how the tight constraints 
on the UCMHs would limit the creation of the PBHs, 
it is worth having a closer look at Fig. \ref{power-total}, 
where we have presented the whole of the constraints on the primordial curvature perturbation from different sectors and at various scales. 
Let us take the scale-dependent $\delta_{min}$ as a reference for a moments. 
The limits are tighter for the constant valued $\delta_{min}$. 
So in order to put less possible constraints on the PBH, it is  enough to just consider the most relaxed one, in this case the scale-dependent $\delta_{min}$. We will see that even this is enough to fully rule them out!
We concentrate on the LIGO range, i.e. $1 M_{\odot} <M_{PBH}< 100 M_{\odot}$. 
In order to have an estimation on how the tight limits coming from the UCMHs would affect PBHs, we use the upper-limit on the amplitude of the curvature power-spectrum and plug this back into Eq. (\ref{beta-initial-PBH}). 
If we further assume a Gaussian statistic for the initial seeds of these objects, 
then we could immediately observe that such a constraint would lead to very tiny value of the initial abundance, $\beta \lll 10^{-10^{6}}$, which is much smaller than necessary to have any contribution on the DM today! 
This has something to do with the very steep decaying behavior of the $erfc{x}$ toward the large values of $x$. 
It can be easily shown that changing the arguments of this function by three order of magnitude leads to the same constraints that we have found above.  Any primordial perturbation PDF that does not have large extra non-gaussian peaks just to make PBHs will have a limit on $f_{DM}$ well below 0.1.
Therefore in order to not produce too many UCMHs, 
consistent with their null observational effects currently, and under the assumption that they both do carry nearly the same initial statistics, we conclude that it is almost impossible to be able to create any PBH within their overlap scales.\\
This constraint is likely only violated if PBHs are made from another source such as in first order phase transitions a la Dolgov and Silk or by collapsing cosmic strings. Both scenarios are unlikely to provide sufficient PBHs in this region.
These constraints are not surprising in a simple slow roll and terminate Inflation model but do provide restrictions 
on the any large bump up in perturbations produced during later time Inflation.
\subsubsection{Pulsar Timing Implications Expanded}
Since  the pulsar timing limits are not sensitive to the type of the DM, we can use them to put new direct constraints on the PBHs as well. In another word, the green dashed-dotted-dotted line for the Pulsar timing could be also interpreted as a tool to put stronger constraints on the fraction of the DM in the form of the PBH. 
This limit is independent of whether the PBHs form from primordial fluctuations, phase transitions or cosmic strings.
The origin is not important as that they exist at the present epoch.\\
So PBHs should be the ideal candidate for the pulsar timing limits
and we interpret the results as independently as limiting  $f_{DM}$ to well less than $10^{-2}$
from $10^{-7} M_\odot \lesssim M_{PBH} \lesssim 10 M_\odot$.
%To limit the window covering $10^{-15} M_\odot  \leq M_{PBH} \leq 10^{-7} M_\odot$,
%we need to use our argument about the limit from primordial perturbations to get below the $f_{DM} \le 10^{-2}$
%for all PBH masses.

%%%%%%%%%%%%%%%%%%%%%%%%%%%%%%%%%%%%%%%%%%%%%%%%%
%%%%%%%%%%%%%%%%%%%%%%%%%%%%%%%%%%%%%%%%%%%%%%%%%

\newpage
\section{Conclusion}
\label{conclusion}
The large scale structure is (usually) thought to arise from the collapse of very tiny density perturbations, ($\sim 10^{-5}$), well after the matter-radiation equality. 
However, larger density fluctuations could seed the structures before or about the epoch of matter-radiation equality. 
There are two (well-known) types of these fluctuations, say PBHs and UCMHs. \\
As they are subjected to an earlier collapse and their initial abundance could be used as a probe of primordial curvature perturbations. 
In addition, since they have a very wide (available) window of wave-numbers, they could shed light about the small-scales which are completely out of the reach of any other cosmological probes. \\
 %%%%%%%%%%%%%%%%%%%%%%%%%%%%%%%%%%%%%%%%%%%%%%%%%
%\begin{adjustwidth}{-0.5cm}{-0.5cm}
\begin{figure}[htbp]
	\centering
	\hspace*{-1.9cm}
	\includegraphics[width=1.14\textwidth]{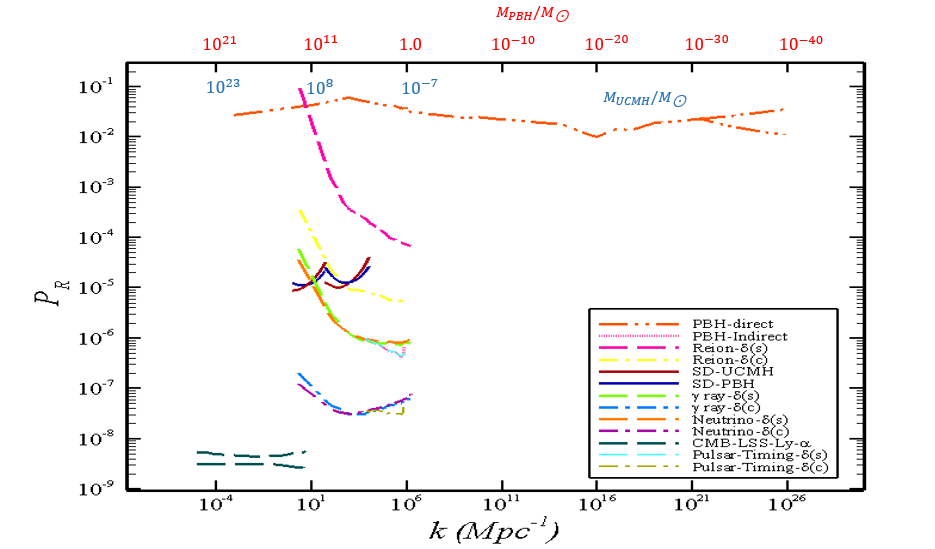}
	\caption{The full set of constraints on the amplitude of the primordial curvature power spectrum. The red dotted-dotted-dashed line denotes the direct constraints on from the PBHs. We also have plotted the indirect constraints on the PBH coming from the UCMHs with purple dotted line. The solid blue and red lines denote the constraints form the CMB spectral distortion. Furthermore, We would also present the limits from the UCMHs. Since the constraints depend on the threshold value of the $\delta$,  we present the limits for both of the constant, denoted by $\delta(c)$, as well as the scale dependent threshold, referred by $\delta(s)$.}
	\label{power-total}
\end{figure}
%\end{adjustwidth} 
\clearpage
%%%%%%%%%%%%%%%%%%%%%%%%%%%%%%%%%%%%%%%%%%%%%%%%%

In this work, we have revisited the bounds on the primordial curvature power-spectrum coming from PBHs and UCMHs. 
Our work and our limits could be summarized in Fig. \ref{power-total}, in which we have calculated and collected the whole, current and futuristic, constraints on the curvature power-spectrum. 
In addition, we have also presented the whole bounds from the combination of the CMB, LSS and the Ly$\alpha$ forest. \\
There are few very important points that are worth mentioning. \\
First of all, the most relaxed kind of constraints  on the primordial curvature spectrum are associated with PBHs, 
which are of order $\mathcal{P}_{R}\sim(10^{-2}-10^{-1})$. 
These limits could be pushed down by several orders of magnitude, $ \mathcal{P}_{R}\sim (10^{-5}-10^{-4})$, 
if we also consider the CMB SD, though.\\
Secondly, the constraints on UCMHs could be divided in two different pieces; those which depend on the nature of the DM such as, $\gamma$-ray, Neutrinos and Reionization. And, those which are blind to the type of the DM within these halos and are only sensitive to the gravitational effects like the pulsar timing. \\
Thirdly, the constraints on UCMHs do also depend on the minimal required value of the initial density fluctuation. There are currently two different choices commonly used for this function; it could be either a constant $\delta_{min} \sim 10^{-3}$ or a function of the scale and redshift. 
Here we reconsidered the constraints for both of these choices to see how much the constraints would change. 
Our calculation showed that, this difference could lead to about $3$-orders of magnitude changes on the upper-bound of the primordial curvature perturbation. 
Indeed, the most severe constraints, $\mathcal{P}_{R}\sim (10^{-7}-10^{-6})$, are coming from the constant valued $\delta_{min}$. \\
Fourthly, except for the relatively relaxed constraints from the Reionization, $\mathcal{P}_{R}\sim (10^{-4}-10^{-1})$, depends on the initial value of $\delta_{min}$ as we pointed it out above, the rest of the bounds on UCMHs are indeed comparable. This is very informative. Because, as we pointed it out before, some of these limits do depend on the nature of the DM as well as their annihilation process etc. Some do not so such a limits are meaningful and independent of the DM details. \\
Fifthly, the constraints from the (current) CMB SD from COBE/FIRAS are comparable to the upper-limits from the scale-dependent $\delta_{min}$. 
However, considering the futuristic types of CMB SD observations could possibly pushed down to be comparable with the constant choice of $\delta_{min}$. \\
Lastly, comparing the constraints from the (very) large scales with that of UCMHs, we could immediately see that they are not too far away. Therefore, although UCMHs might be a bit rare as compared with the usual case, they could give us comparable information in very small scales, down to $k \sim 10^{7} Mpc^{-1}$. This could be thought as an unique way of shedding lights about the fundamental physics at very early universe motivating further investigation of these objects. \\
UCMHs limits basically rule out $\mathcal{O}(10) M_{\odot}$ PBHs as the dark matter
because unless there is a special very non-gaussian bump with about $10^{-20}$ of the area that goes out to $\delta \ge 0.3$
and then nothing down to below $10^{-3}$ the lack of UCMHs rules out PBHs over that long interval
to well below the DM limits.  Then the question is whether there is anything that denies or gets rid of UCMHs
and that appear unlikely.

\section{Acknowledgments}
We are very much grateful to  Andrew Cohen, 
Gary Shiu, Henry Tye and Yi Wang for the helpful discussions. R.E. is grateful to Simon Bird and Martin Schmaltz for the interesting discussions.  
The work of R.E. was supported by Hong Kong University through the CRF Grants of the Government of the Hong Kong SAR under HKUST4/CRF/13. 
GFS acknowledges  the IAS at HKUST and the Laboratoire APC-PCCP, Universit\'{e} Paris Diderot and Sorbonne Paris Cit\'{e} (DXCACHEXGS) and also the financial support of the UnivEarthS Labex program at Sorbonne Paris Cit\'{e} (ANR-10-LABX-0023 and ANR-11-IDEX-0005-02).\\

{}

\end{document}